\documentclass[aps,prb,floatfix,twocolumn,showpacs]{revtex4}
\usepackage{latexsym}
\usepackage{graphicx}
\usepackage{dcolumn}
\newcommand{\figwidth}{3.25 in}
\begin{document}
\title{Quantum simulation of $^3$He impurities 
and of $^4$He interstitials in solid $^4$He}

\author{Keola Wierschem$^{1}$ and Efstratios Manousakis$^{1,2}$}
\affiliation{$^{1}$Department of Physics, 
Florida State University, 
Tallahassee, FL 32306-4350, USA and \\
$^{2}$Department of  Physics, University of Athens,
Penipistimiopolis, Zografos, 157 84 Athens, Greece.}  
\date{\today}
\begin{abstract}
We have studied the role of an atomic $^3$He impurity and an interstitial
$^4$He atom in  two- and three-dimensional solid $^4$He 
using path integral Monte Carlo (PIMC) simulation.
We find that when a substitutional $^3$He impurity is introduced, the impurity 
becomes localized and occupies an ideal lattice site.
When an interstitial  $^3$He impurity is introduced in the $^4$He solid,
we find that the impurity becomes localized at a substitutional 
position and, thus, promotes the extra $^4$He atom to the 
interstitial space. As a consequence we find that
the one-body density matrix (OBDM)  and the superfluid fraction,
for the case of a $^4$He solid
with an interstitial impurity, are very similar to those calculated
for a $^4$He solid with a $^4$He interstitial atom.
Namely, while the off-diagonal OBDM  approaches zero exponentially 
with increasing 
particle displacement for the ``pure'' solid, an interstitial $^4$He atom or
a $^3$He  impurity appear to enhance it at 
long distances. Finally, the effective mass of the $^3$He impurity 
quasiparticle in 2D and 3D crystalline $^4$He is estimated.
\end{abstract}
\maketitle

\section{Introduction}
The torsional oscillator experiments of Kim and Chan\cite{chan}, where
at low temperature a drop in the moment of inertia is observed, have
motivated a number of  computational 
studies\cite{prokofev,ceperley,mas_triag,worm,Bon1,Pollet,Bon2,
pressure} of solid 
$^4$He as well as various theoretical proposals\cite{Dorsey,manousakis,Toner} 
to explain the observation. 

There is evidence of a very strong dependence of the
superfluid response on the $^3$He impurity concentration\cite{chan3} 
as well as other well known facts\cite{goodkind} about the role 
of impurities in solid $^4$He\cite{smith,rudavskii,ganshin}.
Proposals for the possible role of $^3$He impurities in solid $^4$He 
have a long history and date back in the late 
sixties\cite{andreev} and seventies\cite{andreev2,kagan}. In addition,
there are several experimental studies of the NMR relaxation
of such impurities in solid helium\cite{nmr}. It is, therefore, of
great interest to study the role of impurities in solid $^4$He.
Boninsegni et al.\cite{Bon1} have carried out a 
path integral Monte Carlo (PIMC) simulation of 
three-dimensional (3D) solid $^4$He using the worm algorithm and found
that vacancies phase separate. Pollet et al.\cite{Pollet} and Boninsegni et
al.\cite{Bon2} have also used the above PIMC technique to show that grain
boundaries in solid $^4$He and screw dislocations lead to superfluidity.
In addition, using the same method Pollet et al.\cite{pressure} have 
shown that the gap to create vacancies  closes by applying a moderate pressure.

In the present paper, motivated by the recent experimental and 
theoretical activity on the possible role of the $^3$He impurities 
in solid $^4$He, we study the role of a $^3$He impurity and of an interstitial
$^4$He atom  in two-dimensional (2D) and 3D solid $^4$He using 
PIMC  simulation. In addition to the 
motivation generated by the previous discussed experimental activity,
this problem is of interest in its own right because it is not really known 
what happens locally when one injects a $^3$He atom  in 2D or 3D solid $^4$He,
and  this can be studied by quantum simulation.
In particular, we use the worm
algorithm~\cite{worm} to simulate the 2D and 3D
solid helium in the presence of such crystalline defects.
We present results of the radial distribution functions and off-diagonal
one-body density matrix (OBDM) for the following cases.
(a) Pure solid $^4$He at somewhat above but near the liquid-solid melting 
density ($\sigma=0.070$ \AA$^{-2}$ or $\sigma=0.026$ \AA$^{-3}$).
(b) A single substitutional $^3$He impurity in solid $^4$He.
(c) An interstitial $^4$He atom (defect). This atom is identical to the
other $^4$He atoms and, therefore, it participates in permutation cycles.
(d) An interstitial $^3$He impurity in solid $^4$He. 

We find that an initial interstitial impurity quickly relaxes
to a regular lattice site of the $^4$He solid by creating an
interstitial $^4$He atom as was proposed in Ref.~\onlinecite{manousakis}.
Furthermore, we find that introducing such interstitial impurities
in $^4$He solid greatly enhances the long-distance part of the
off-diagonal OBDM. This enhancement as well as the calculated 
superfluid response is comparable to that of interstitial 
$^4$He atomic defects. It is quite possible that at a finite density
interstitial $^4$He atoms phase separate as do vacancies\cite{Bon1}. In 
such case the enhancement of the OBDM at long distances and of the superfluid
density, due to a single interstitial $^3$He or $^4$He atom
found in the present paper, may disappear when a finite density of
such impurities or interstitials is introduced. However, interstitial
atoms or impurity atoms can bind to already existing defects, such as 
dislocations or disclinations (especially in 2D) and this tendency for phase
separation may be avoided. In general, it is of great value to know what
happens {\it locally} in the 2D and in the 3D crystalline $^4$He 
when a $^3$He impurity or an interstitial $^4$He atom is introduced.

The paper is organized as follows. In Sec.~\ref{method} we briefly
describe the PIMC method used to study this system.
In Sec.~\ref{distribution}, we present and discuss the pair distribution
functions for the case of a 2D and 3D solid with and without
the introduction of a $^3$He impurity and $^4$He interstitial atom.
The energetics
of creating such atomic defects in the 2D and 3D solids as well
as the calculation of the effective mass of $^3$He impurity in solid $^4$He 
is discussed in Sec.~\ref{energetics}.
In Sec.~\ref{superfluidity} we present the results 
for the one-body density matrix, the superfluid density and a histogram
of the number of particles involved in the same permutation cycle for the
cases (a-d) above for 2D and 3D solid helium.
Finally, the main findings  as well as
the limitations of the present work are discussed in Sec.~\ref{discussion}. 
\section{Simulation details}
\label{method}
Using an approximation for the density matrix that is accurate to fourth
order~\cite{cuervo} in $\tau$, we use 320 time slices to reach a
simulation temperature of $1 K$. We have collected data from 2500
continuous iterations for our simulations in 2D, and $\sim$1000
continuous iterations for our simulations in 3D.
Each iteration consists of 500 Monte Carlo moves.

All simulated atoms considered in our present studies are
isotopes of helium  and therefore interact via the same potential. 
We use the Aziz~\cite{aziz} potential to model both the
$^4$He-$^4$He interaction and the $^3$He-$^4$He interaction.
With the exception of the $^3$He impurity atom
the rest are all $^4$He atoms which will be treated appropriately
to simulate their bosonic nature. The impurity atom
is distinguishable from the ``background'' $^4$He atoms.

Our simulation cell is designed to accommodate either a 2D 56-site
triangular lattice that is very nearly square
($25.86$~\AA~$\times~25.60$~\AA), or a 3D 180-site
hexagonal close-packed lattice
($18.35$~\AA~$\times~19.07$~\AA$\times~17.98$~\AA).
In both cases of the 2D and 3D lattices we have used periodic boundary 
conditions.
The density of lattice sites is
fixed at $0.0846 $\AA$^{-2}$ (2D) and $0.0286 $\AA$^{-3}$ (3D).
We will use the term {\it pure solid} for the case where there is
exactly one $^4$He atom per lattice site. The term
{\it substitutional solid} will be used when a single $^4$He atom
is removed from the pure solid and is replaced with an impurity atom.
Additionally, the term {\it interstitial solid} will be used when a
single atom (either $^4$He or an impurity) is added to the pure solid.

\section{Distribution functions}
\label{distribution}
\subsection{Two-dimensional solid $^4$He}
\begin{figure}[htp]
\vskip 0.2 in
\begin{center}
\includegraphics[angle=360,width=\figwidth]{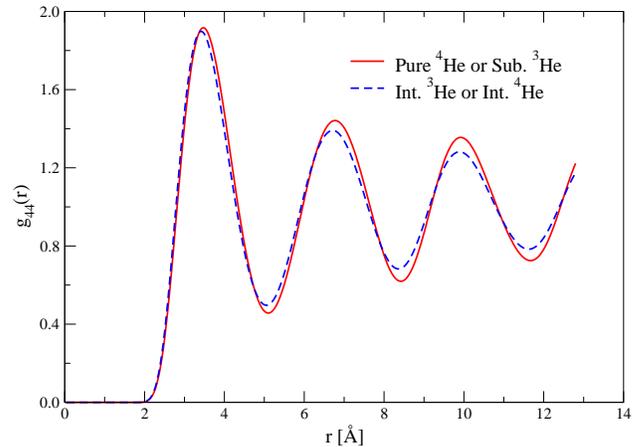}
\end{center}
\caption{Here we show the radial distribution function, $g_{44}(r)$,
for pairs of $^4$He atoms in two dimensions. The organizational
structure of the $^4$He atoms does not change in the presence of a
substitutional impurity. However, when an interstitial defect or
impurity is present, we can see that $g_{44}(r)$ becomes less
peaked at the nearest-neighbor distance lattice positions.}
\label{fig1}
\end{figure}
\begin{figure}[htp]
\begin{center}
\includegraphics[width=\figwidth]{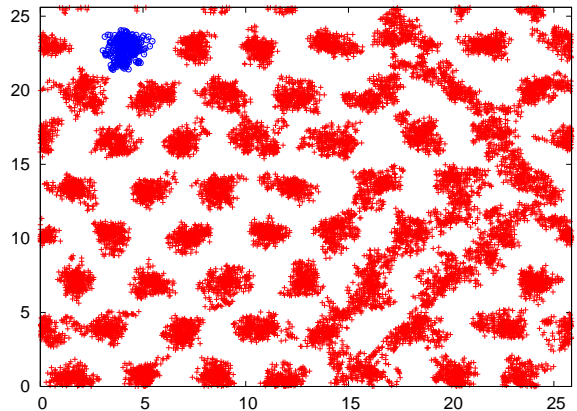}
\end{center}
\caption{A snapshot of a space-time configuration for the 2D
triangular solid, after thermalization, and starting from a
configuration with an interstitial $^3$He atom. Each atom's trajectory
in imaginary time appears as fractal covering a finite size spot.
The crosses (red in the online version) are the $^4$He atoms and the 
circles (blue in the online version) is the $^3$He impurity atom. }
\label{fig2}
\end{figure}
\begin{figure}[htp]
\begin{center}
\includegraphics[width=2.5 in ]{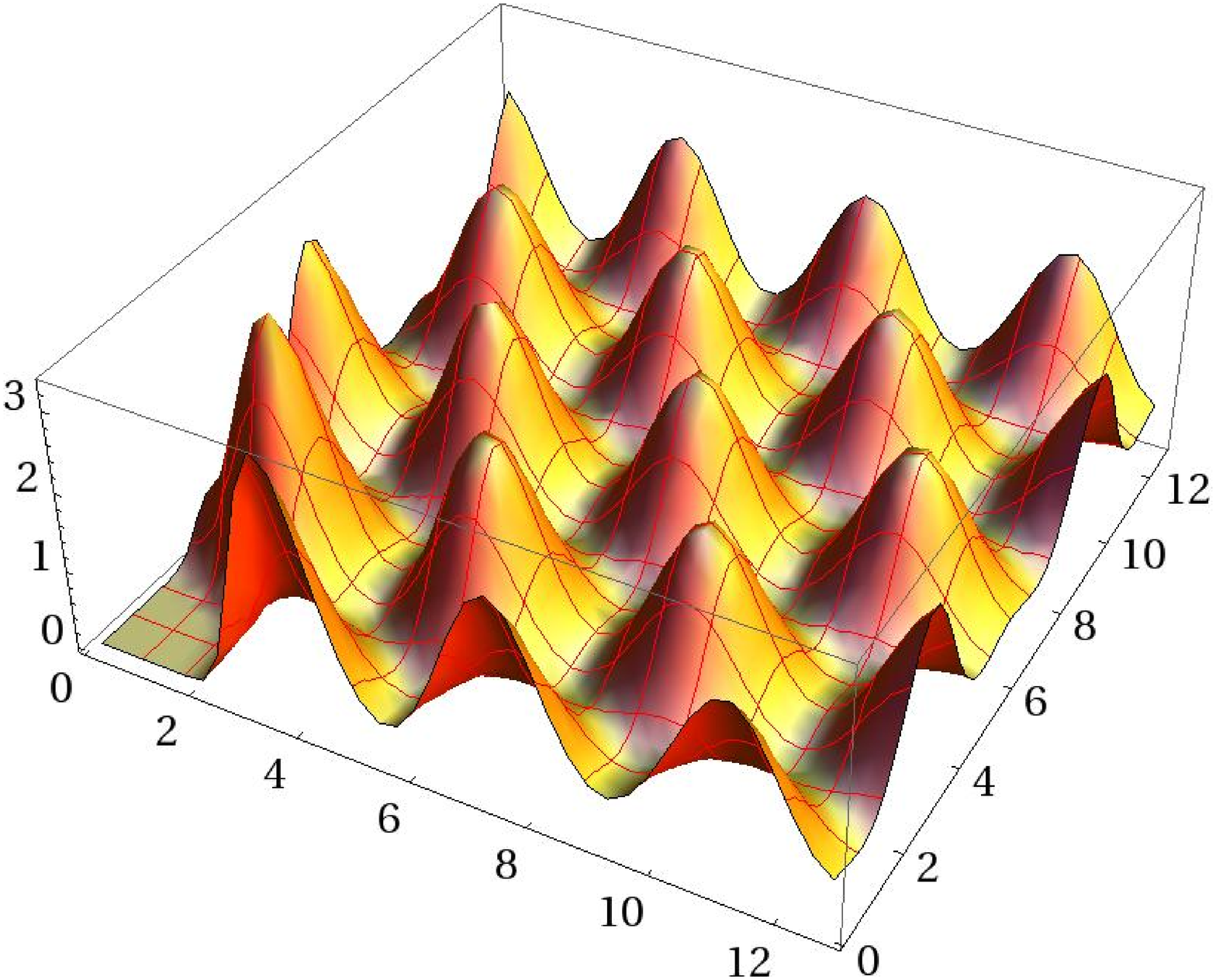}
\includegraphics[width=2.5 in]{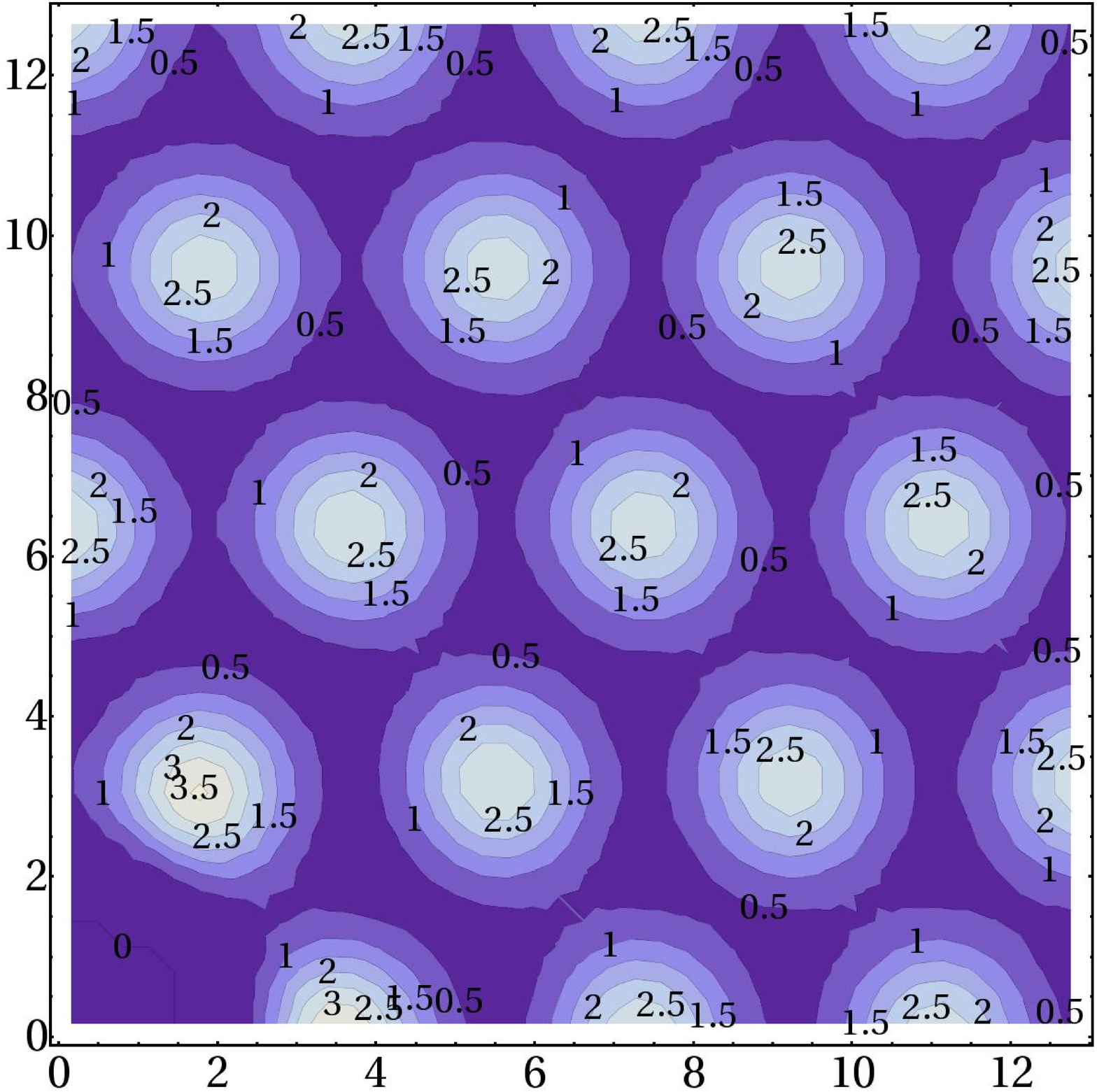}
\includegraphics[width=\figwidth]{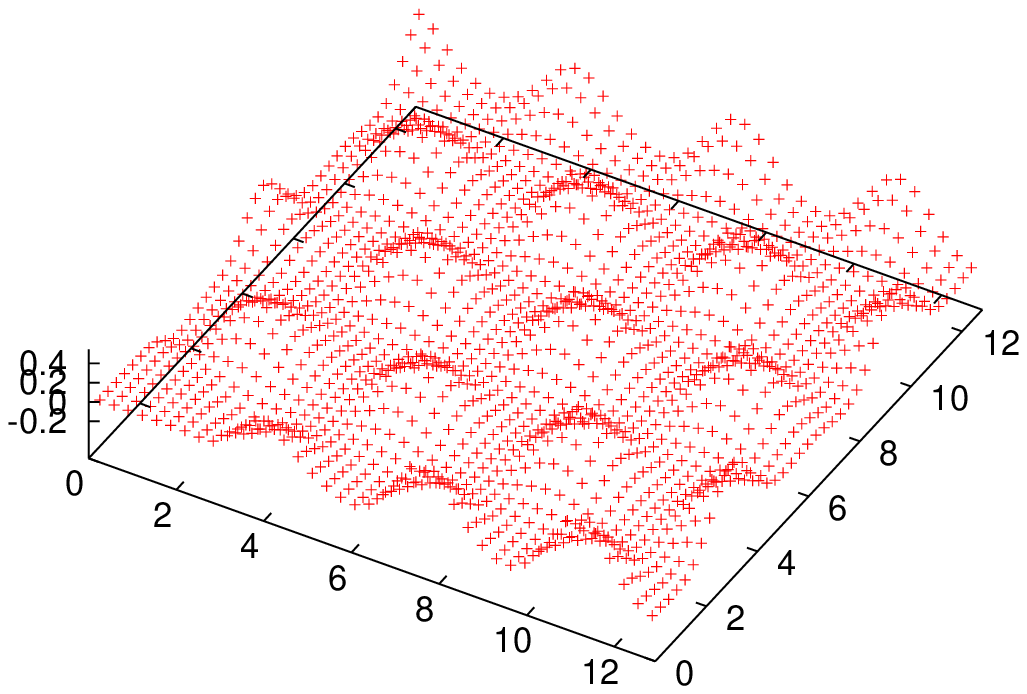}
\caption{Top: The pair distribution function $g_{44}(x,y)$ for pure 2D solid 
$^4$He. Middle: Contour plot of the distribution function for pairs of 
$^4$He atoms, $g_{44}(x,y)$ in the pure solid.
Bottom: the difference $\delta g_{44}$ between $g_{44}$ of the pure
(solid) and that of the solid with interstitial impurity.}
\label{fig3}
\end{center}
\end{figure}
\begin{figure}[htp]
\begin{center}
\includegraphics[width=2.5 in]{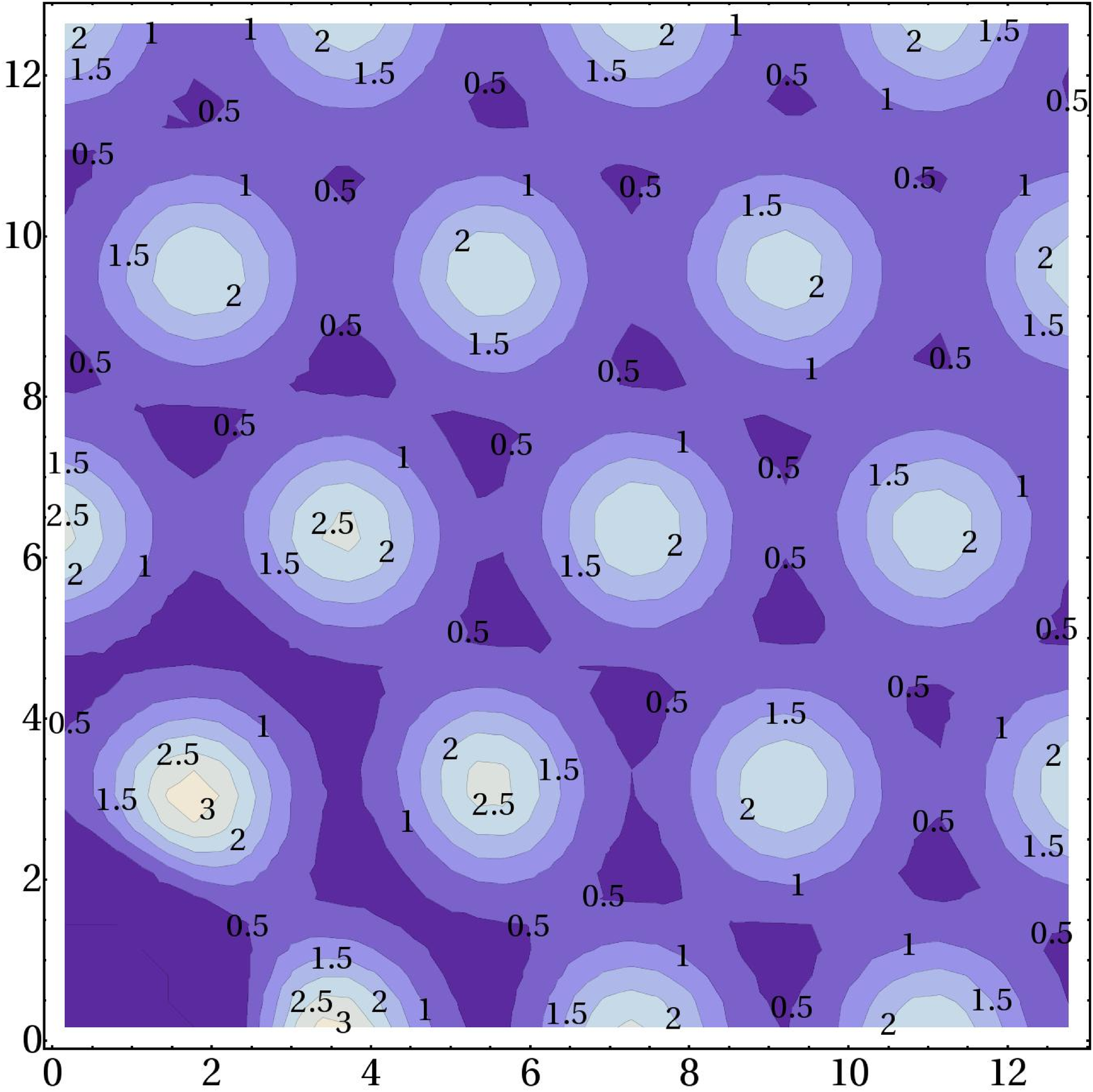}
\includegraphics[width=2.5 in]{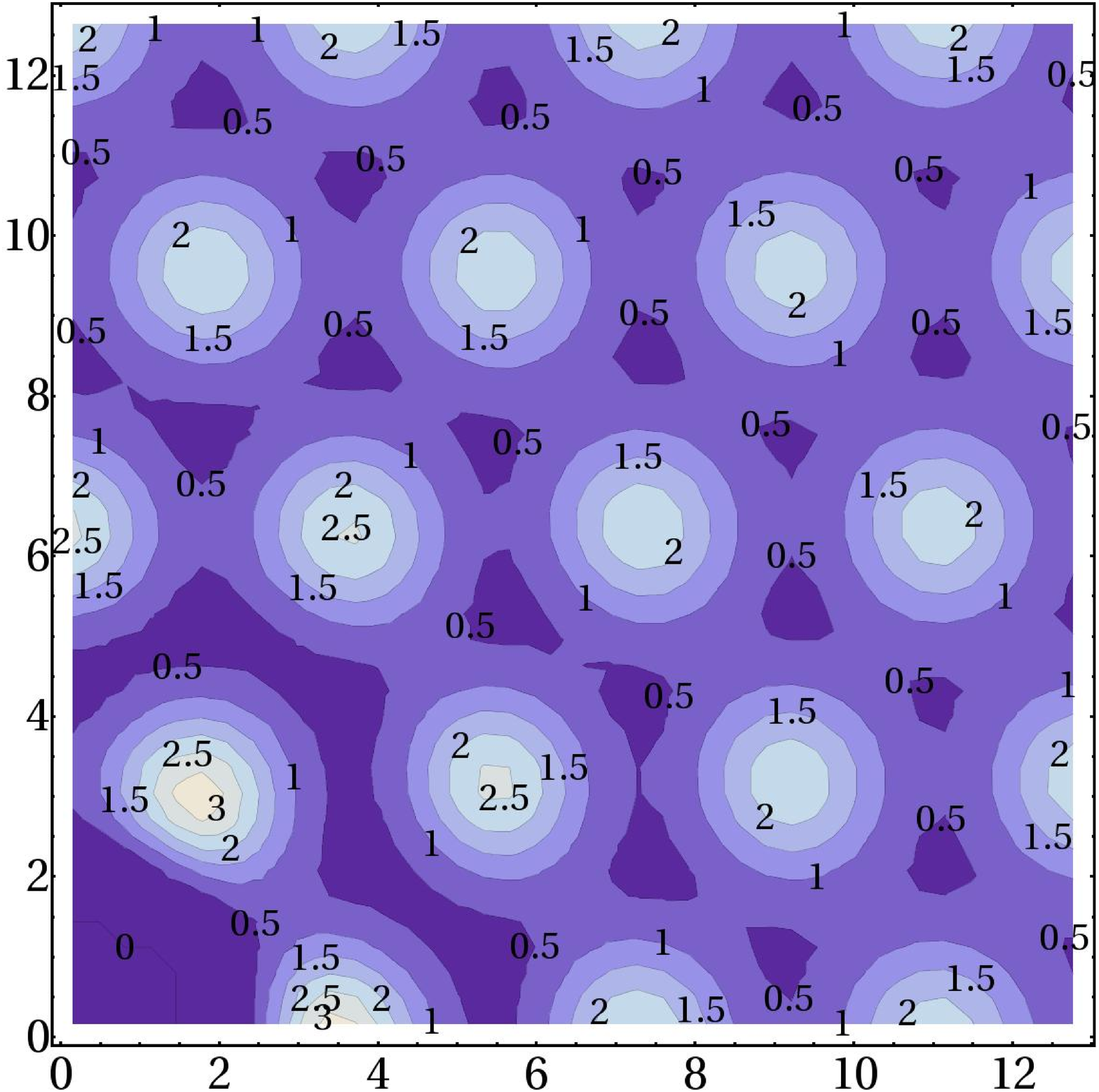}
\end{center}
\caption{Top panel: Contour plot $g_{44}(x,y)$ for
the interstitial solid. This function is independent
of the type of defect or impurity.
Bottom panel: Contour plot of the
distribution function  $g_{34}(x,y)$ for the same interstitial solid. }
\label{fig4}
\end{figure}
\begin{figure}[htp]
\begin{center}
\includegraphics[angle=360,width=\figwidth]{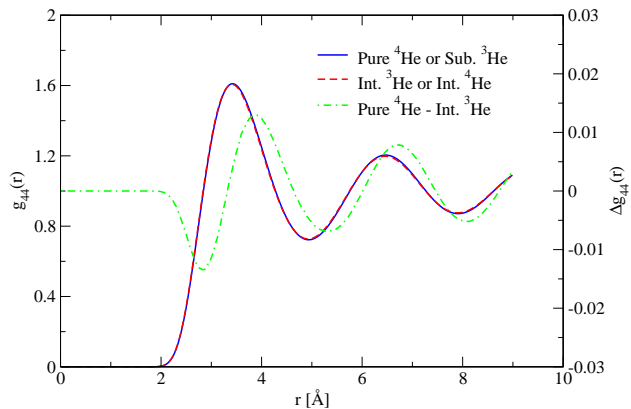}
\end{center}
\caption{The radial distribution function for pairs of $^4$He atoms
in the three-dimensional HCP lattice simulation cell, $g_{44}(r)$.
The organizational structure of the $^4$He atoms does not change
in the presence of a substitutional impurity.
However, when an interstitial defect or impurity is present,
by looking at $\Delta g_{44}(r)$ (scale on the right) 
we can see that $g_{44}(r)$ becomes
less peaked at the nearest-neighbor distance lattice positions.}
\label{fig5}
\end{figure}

How does the impurity atom affect the pair distribution function
$g_{44}$ of the $^4$He atoms of the underlying solid?
We find that when a substitutional impurity is introduced it becomes
localized and occupies an ideal lattice position with its own
zero-point motion determined by its different mass.
In Fig.~\ref{fig1} we present the calculated $g_{44}(r)$ radial 
distribution for pairs of $^4$He atoms for the four different case
systems studied:
(a) pure solid $^4$He (dashed line)
(b) the $^4$He solid with a substitutional $^3$He impurity (also dashed line)
(c) the $^4$He solid with an interstitial $^4$He defect (solid line), and
(d) the $^4$He solid with an interstitial $^3$He impurity (also solid line).
Within the accuracy of our results we cannot discern 
any difference in the $g_{44}$ distribution function for the cases of
the pure solid and the substitutional impurities. When an
interstitial impurity is present in the $^4$He solid, we find that
the impurity becomes localized at a substitutional position,
thereby promoting the extra $^4$He atom to the interstitial band.
This is shown by the snapshot space-time configuration
shown in Fig.~\ref{fig2}. Notice that while the initial configuration
has an interstitial $^3$He impurity, in the configuration obtained 
after thermalization (shown in Fig.~\ref{fig2})  the $^3$He becomes
substitutional by promoting an interstitial $^4$He atom. Namely, in the
equilibrium configuration, shown in Fig.~\ref{fig2},
the $^3$He atom, in our lattice with periodic boundary conditions,
is located in a regular triangular lattice position surrounded by six
$^4$He atoms. In addition, a $^4$He atom has been promoted to the
interstitial space which creates larger density fluctuations in 
the crystalline arrangement in some parts of the system.
As a consequence of this fact $g_{44}(r)$,
in Fig.~\ref{fig3}, is less peaked at the lattice positions.  
In Fig.~\ref{fig3} (top) the calculated pair distribution function 
$g_{44}(x,y)$  for pure 2D solid $^4$He is shown and in
Fig.~\ref{fig3} (middle) 
we present the contour plot of the same $g_{44}(x,y)$.
This function is nearly identical for the substitutional solid
(which is not shown, as it looks exactly alike). This implies that
the introduced substitutional impurity becomes localized and
it only affects its neighboring atoms. In the case of an interstitial
impurity  the difference in the $g_{44}$ distribution function,
as discussed above and shown in Fig.~\ref{fig1} and Fig.~\ref{fig3}
(bottom), 
is significant because the added impurity takes the position of
a $^4$He atom and, thus, there is an extra $^4$He atom that necessarily
becomes interstitial. In the bottom panel of Fig.~\ref{fig3}
we plot $\delta g_{44}(x,y)$, the difference
between $g_{44}(x,y)$ of the pure solid and the solid with a single 
interstitial impurity. Notice that the extra atom is
truly interstitial since the $g_{44}$ is reduced by an amount
in the neighborhood of the ideal lattice positions and enhanced in the
interstitial space by the same amount. 
It was verified through integration in the enhanced regions (or the reduced
regions) finding exactly one extra $^4$He atom in the interstitial regions.

Our finding that the interstitial impurity becomes localized at 
regular lattice sites can be further illustrated by 
comparing the contour plots of the $g_{44}(x,y)$ and $g_{34}(x,y)$ 
for the case where we have a $^4$He solid with an interstitial impurity.
In the top panel of Fig.~\ref{fig4}, we present the contour plot of
the distribution function $g_{44}(x,y)$ for the case of a $^4$He solid with
an interstitial impurity. Within the accuracy of the discretization of
the probability density of the contour plot this function is independent
of the type of defect or impurity.
In the lower panel is the
distribution function $g_{34}(x,y)$ for pairs consisting of the impurity atom
and one $^4$He atom. 
Because the contour plots for both $g_{34}(x,y)$ and $g_{44}(x,y)$ 
are identical in shape and in
form, we may surmise that the impurity atoms are located at
lattice sites.

\subsection{Three-dimensional solid $^4He$}
In Fig.~\ref{fig5} we show $g_{44}(r)$ for the 3D system. As in 2D,
we find that the pure solid and the substitutional solid are nearly
identical in structure, as are the two interstitial solids. Also shown
is the difference, $\Delta g_{44}(r)$, between $g_{44}(r)$ the pure
solid and the interstitial solid. As expected, $g_{44}(r)$ for both
interstitial solids is less peaked at lattice positions compared to
the pure and substitutional solids. This indicates that the $^4$He
interstitial solid really does have a $^4$He atom in the interstitial
space, and also that the interstitial $^3$He solid has relaxed into
a space where the $^3$He interstitial atom has become
substitutional, and in doing so promoted a $^4$He atom to the
interstitial band.

\begin{figure}[htp]
\vskip 0.3 in
\begin{center}
\includegraphics[width=\figwidth]{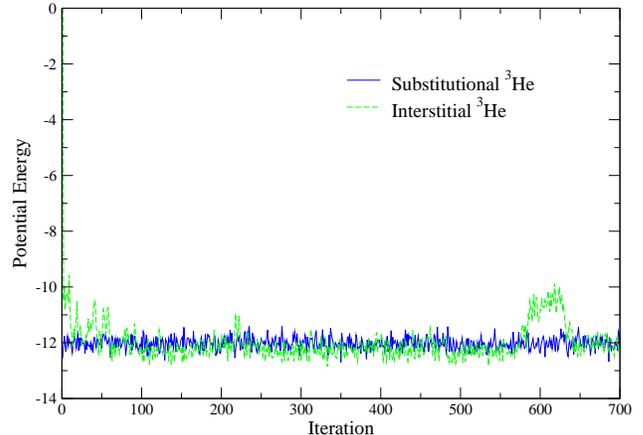}
\end{center}
\caption{Potential energy of a $^3$He atom placed either
substitutionally solid line (blue in the online version) 
or interstitially dashed line (green in the online version)
into triangular solid $^4$He. After a brief relaxation,
both energy values remain close except for occasional ``blips''
in the potential energy of the initially interstitial $^3$He atom.}
\label{fig6}
\vskip 0.1 in
\end{figure}

\begin{figure}[htp]
\begin{center}
\includegraphics[width=\figwidth]{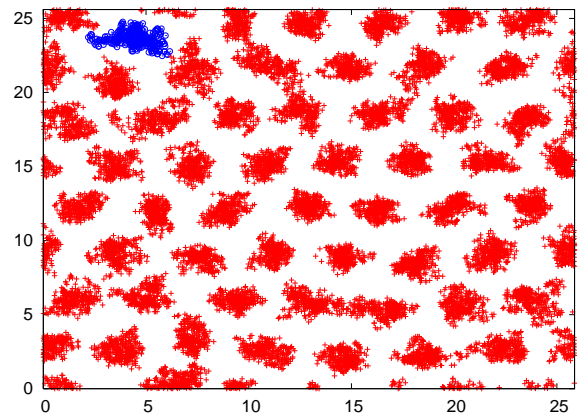}
\end{center}
\caption{Snapshot of a space-time configuration of the $^3$He
interstitial solid in a ``blip'' of elevated potential energy
that appears after thermalization. Red crosses represent the
$^4$He atoms at each imaginary time slice, while blue circles
represent the $^3$He impurity. The $^3$He atom can be seen to
be at a region of local disorder.}
\label{fig7}
\end{figure}
\section{Energetics of impurity and interstitial}
\label{energetics}
If a $^3$He atom, initially placed in the interstitial region of
a triangular solid of $^4$He atoms, relaxes onto a lattice site
by promotion of a $^4$He atom to the interstitial space, this
should be seen in the energy values of the simulated atoms.
In Fig.~\ref{fig6} we show the potential energy of
a $^3$He atom in the substitutional and interstitial $^3$He solids.
A short relaxation time can be seen for the interstitial solid,
as the $^3$He atom relaxes onto the lattice. After that,
the potential energy of a $^3$He atom in both systems is almost
the same. After 600 iterations, a small bump is seen in the energy
of the (initially) interstitial $^3$He atom. A snapshot of the
atomic configuration at this elevated energy value is shown in
Fig.~\ref{fig7}. The $^3$He atom is no longer at an
equilibrium lattice position, but rather at what appears to be a
possible edge dislocation. This is not entirely unexpected, as a
$^3$He atom in solid $^4$He exhibits a high rate of diffusion.
Such ``blips'' in the energy of the $^3$He in the interstitial solid
occur occasionally throughout our simulation, but account for no more
than 5\% of configurations.

\begin{table}[hb]
\begin{tabular}{ccc}
                          &         2D         &        3D        \\
Int. $^4$He - Pure $^4$He & 50.27 $\pm$ 0.54 K & 22.4 $\pm$ 1.3 K \\
Int. $^3$He - Sub. $^3$He & 50.41 $\pm$ 0.55 K & 24.1 $\pm$ 1.2 K \\
\end{tabular}
\caption{Excitation energy of an interstitial $^4$He atom, as calculated
by the difference in energy between
(1) the pure solid and the interstitial $^4$He solid, and
(2) the substitutional solid and the interstitial $^3$He solid.}
\label{table0}
\end{table}

\begin{figure}[htp]
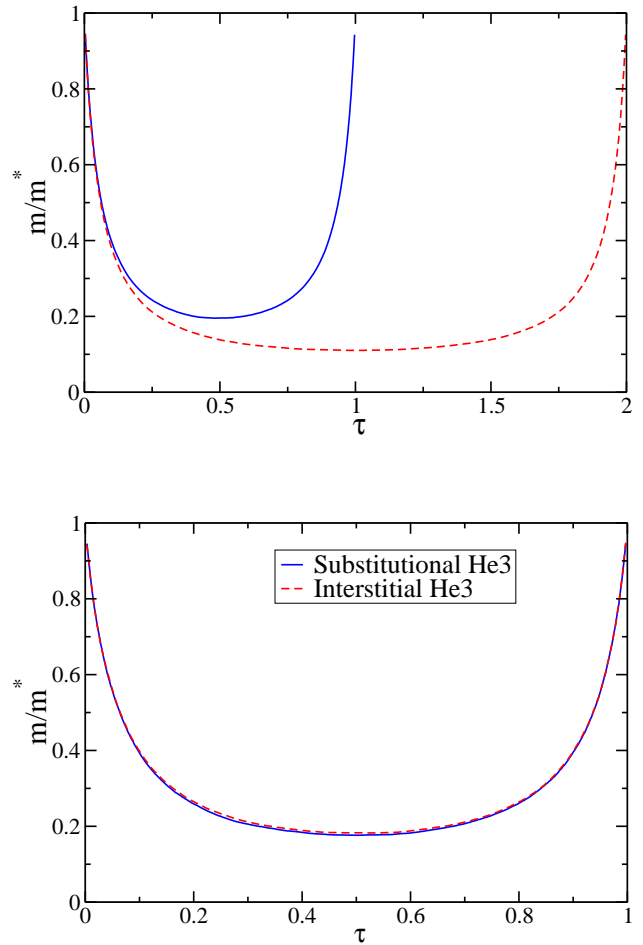

\vskip 0.2 in
\begin{center}
\includegraphics[angle=360,width=\figwidth]{Fig8a.eps}
\vskip 0.4 in
\includegraphics[angle=360,width=\figwidth]{Fig8b.eps}
\end{center}
\caption{Top: The ratio $m/m^*$ as a function of $\tau$ for 2D and 
$T=1$ K (Solid line) and
$T=0.5 $K (dashed line). 
Bottom: The ratio $m/m^*$ as a function of $\tau$ for 3D and $T=1$ K 
for substitutional and interstitial $^3$He impurities.}
\label{fig8}
\vskip 0.2 in
\end{figure}

In Table~\ref{table0} we show the activation energy of an interstitial
$^4$He atomic defect. This is calculated by subtracting the total energy
of the pure solid from the total energy of the interstitial 
$^4$He solid. If the interstitial $^3$He solid is actually the
substitutional solid with an added interstitial $^4$He atom,
as we propose it is based on the distribution functions above,
then the activation energy can also be calculated by subtracting the
total energy of the substitutional solid from the total energy of the
interstitial $^3$He solid. We find that both methods give activation
energies in agreement with one another.

We have also estimated the effective mass of the $^3$He impurity
in solid 2D and 3D $^4$He using our data on the imaginary time diffusion 
following Ref.~\onlinecite{BC}. Namely, we approximate the low-energy
(which dominates the long time evolution) impurity quasiparticle spectrum by
the dispersion near the $\Gamma$ point of the Brillouin zone of both
the triangular 2D solid and of the hexagonal closed packed 3D lattice
\begin{eqnarray}
E(k)=\Delta + {{\hbar^2 k^2} \over {2m^*}}.
\end{eqnarray}
It is straightforward to carry out the imaginary-time evolution 
for this spectrum and to calculate the average of 
$({\bf r}(0)-{\bf r}(\tau))^2$, where ${\bf r}(\tau)$ is 
the impurity coordinate in imaginary time.
We find that,
\begin{eqnarray}
{m \over {m^*}} = \lim_{\tau \to \beta/2} 
 {{\langle ({\bf r}(0)-{\bf r}(\tau))^2 \rangle}
\over { 2 d \lambda} }
{{\beta} \over {\tau (\beta-\tau)}},
\end{eqnarray}
where $\lambda=\hbar^2 / (2 m)$ and $d$ is the dimensionality.
In Fig.~\ref{fig8} we plot the right-hand-side of the above 
equation as calculated from our simulation for the 2D (Fig.~\ref{fig8}(top))
 and 3D (Fig.~\ref{fig8}(bottom)) case.
We find that in the 2D case the effective mass ratio of the $^3$He impurity
at $T=1 K$ is $5.10\pm0.02$  while at $T=0.5$ K it increases
to    $9.06\pm 0.04$. In the 3D case we have available results only for
$T=1$ K, where the substitutional and the interstitial impurity masses 
are found to be $5.67 \pm 0.03$ and $5.47 \pm 0.04$ respectively. 

\begin{figure}[htp]
\vskip 0.2 in
\begin{center}
\includegraphics[angle=360,width=\figwidth]{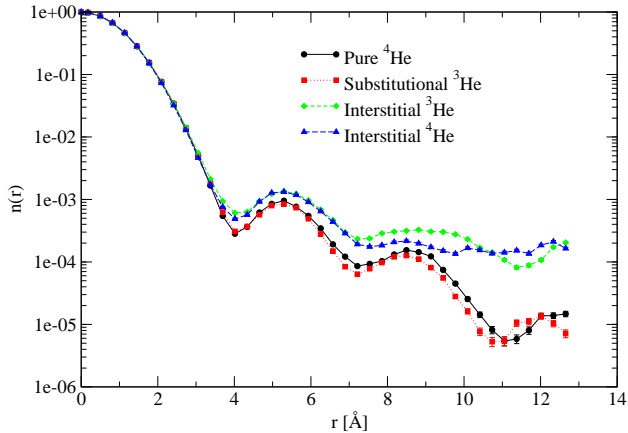}
\end{center}
\caption{The one-body density matrix, $n(r)$. Although no difference is
observed between the pure solid and the substitutional solid,
the interstitial solid clearly shows a significant enhancement
of $n(r)$ quantity.}
\label{fig9}
\vskip 0.2 in
\end{figure}
\section{Off-diagonal one-body density matrix}
\label{superfluidity}
\begin{figure}[htp]
\vskip 0.1 in
\begin{center}
\includegraphics[angle=360,width=\figwidth]{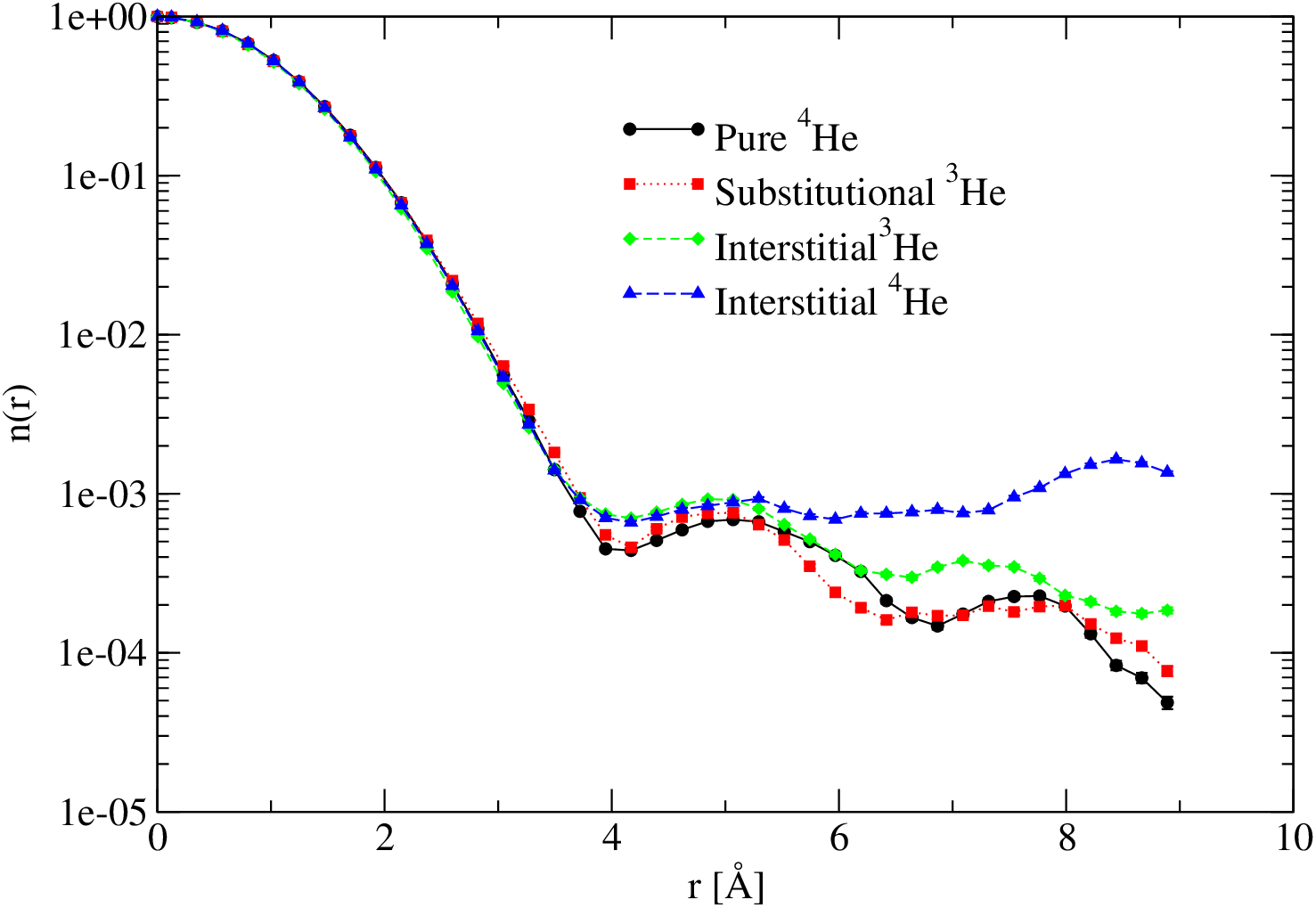}
\end{center}
\caption{The one-body density matrix, $n(r)$, of the 180-site HCP
system in three dimensions. Although no difference is
observed between the pure solid and the substitutional solid,
the interstitial solid clearly shows a significant enhancement
of $n(r)$ quantity.}
\label{fig10}
\end{figure}
\begin{figure}[htp]
\vskip 0.2 in
\begin{center}
\includegraphics[angle=360,width=\figwidth]{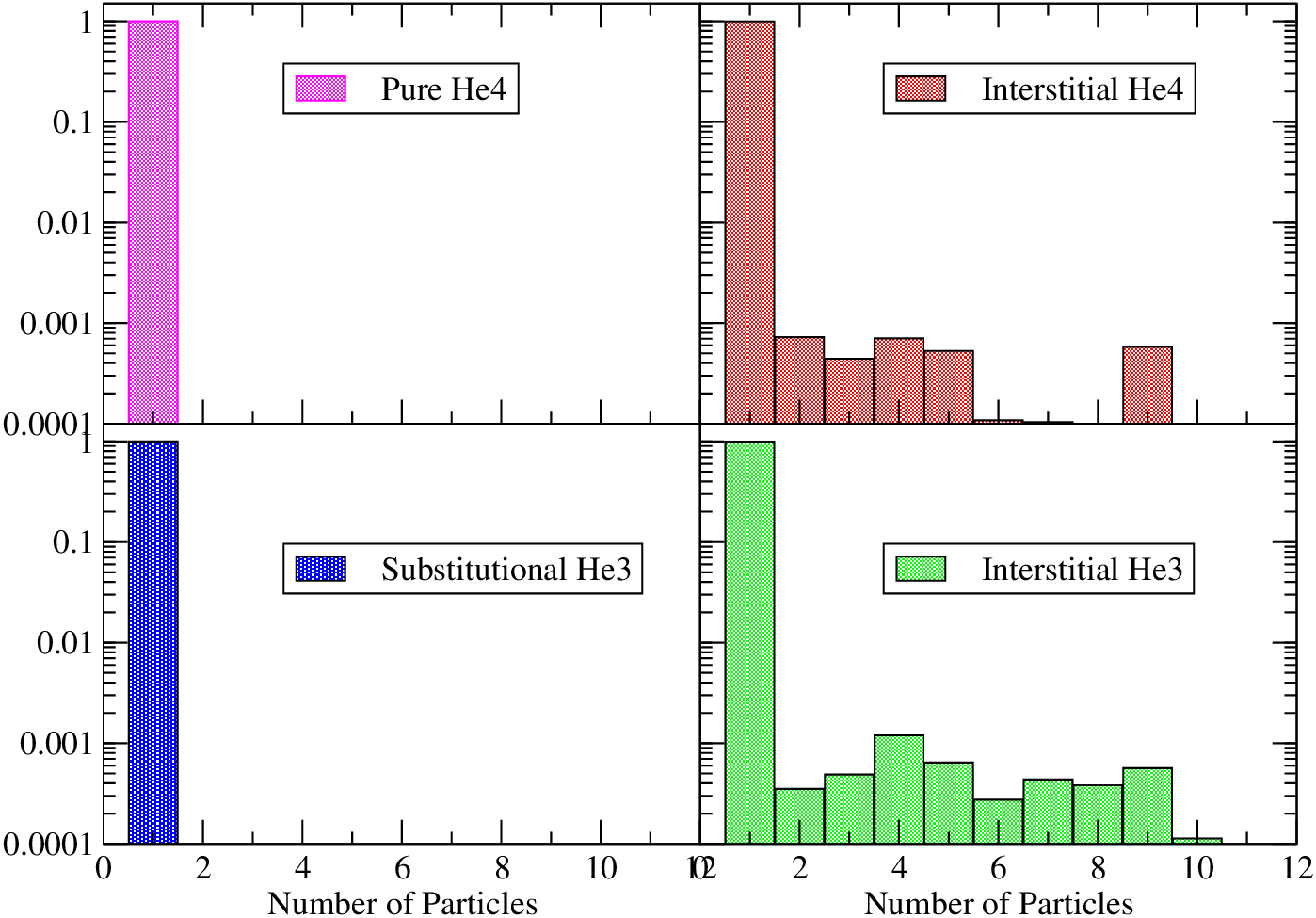}
\end{center}
\caption{Histogram of relative frequency of accepted particle 
permutations for various number of particles in 2D.}
\label{fig11}
\vskip 0.1 in
\end{figure}
\begin{figure}[htp]
\vskip 0.2 in
\begin{center}
\includegraphics[angle=360,width=\figwidth]{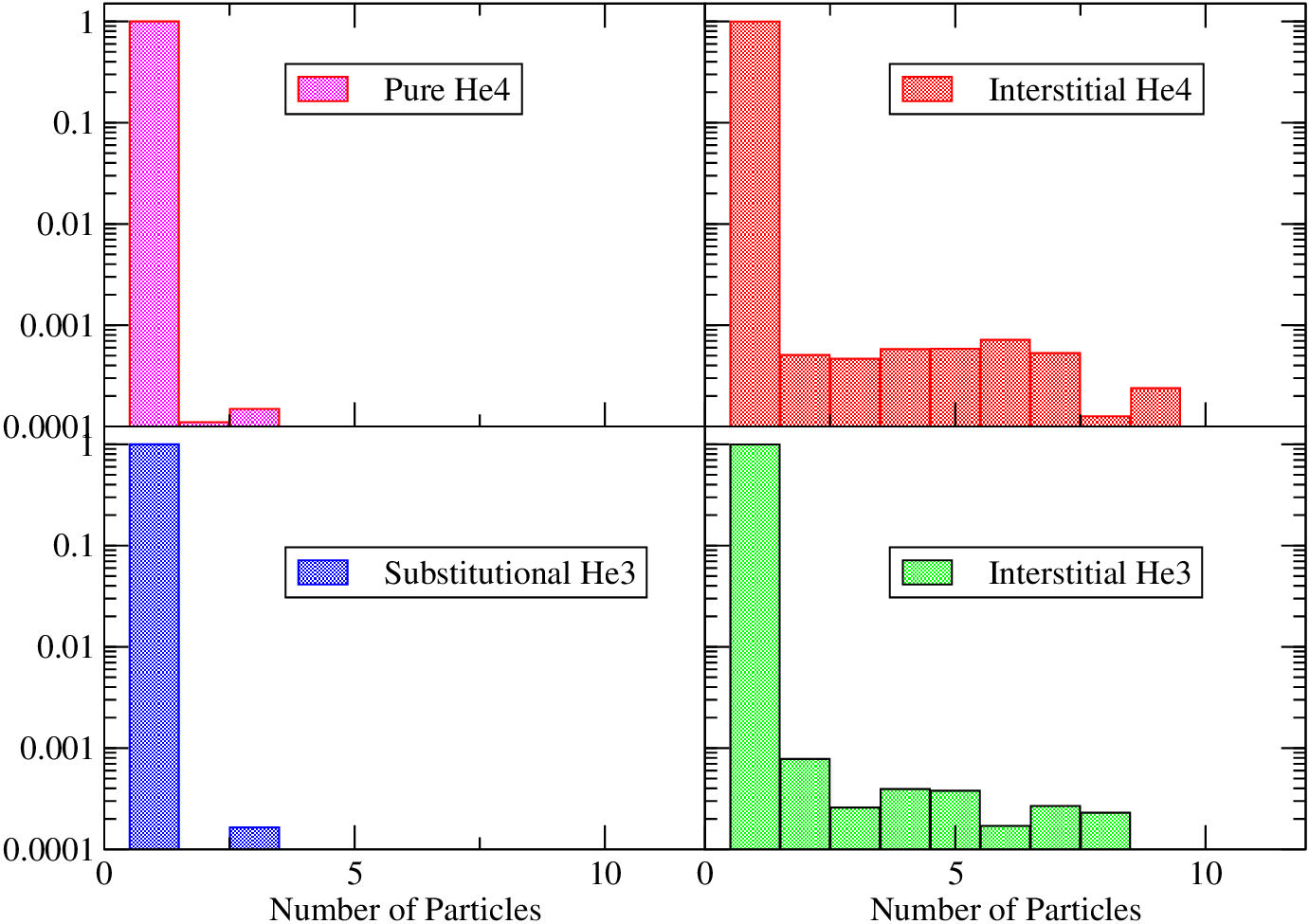}
\end{center}
\caption{Histogram of relative frequency of accepted particle 
permutations for various number of particles in 3D.}
\label{fig12}
\vskip 0.1 in
\end{figure}

\begin{table}[hb]
\begin{tabular}{ccc}
                    &   2D     &   3D     \\
Interstitial $^3$He & 0.021(7) & 0.007(4) \\
Interstitial $^4$He & 0.011(6) & 0.012(5) \\
\end{tabular}
\caption{Supersolid fraction, $\rho_s/\rho$, in the presence of an
interstitial atom. No global permutations were observed for the
perfect lattice and the substitutional impurity.}
\label{table1}
\end{table}


In Fig.~\ref{fig9} we compare the one-body density matrix $n(r)$ for 
(a) defect-free solid $^4$He (solid line),
(b) solid $^4$He with a substitutional $^3$He impurity (dotted line),
(c) solid $^4$He  with an interstitial $^4$He defect (long-dashed line),
(d) solid $^4$He with an interstitial $^3$He impurity (dashed line), and
Notice that the substitutional $^3$He impurity and the pure solid
have similar one-body density matrices. On the contrary,
a $^4$He solid with interstitial $^3$He impurity and
a $^4$He solid with interstitial $^4$He atoms have one-body density
matrices which are significantly enhanced at long distances.
This result agrees with the fact that
winding numbers (and hence superflow) are observed in the
interstitial solid (see Table~\ref{table1}). Notice that
these superfluid fractions are very high considering that the
simulation was carried out at $1 K$. The reason for these 
high superfluid fractions is finite size effects. These results 
for the superfluid fraction are presented in order to make the
case that a interstitial impurity has a very similar effect
on the superfluid fraction and OBDM as an interstitial $^4$He atom.

In Fig.~\ref{fig10} we compare the one-body density matrix for 
the 3D results. As in 2D, both the pure solid and the substitutional
solid show exponential decay of $n(r)$. Although the enhancement of
$n(r)$ at large distance is not obvious for the interstitial $^3$He
solid, it is very obvious for the interstitial $^4$He solid. This may
be due to a shorter MC run as compared to the 2D data. In any case,
once again both interstitial solids display superfluidity, while the
pure and substitutional solids do not (see Table~\ref{table1}).

In Fig.~\ref{fig11} (for 2D) and in Fig.~\ref{fig12} (for 3D) we present 
a histogram of cycles (i.e., how often in the simulation 
we encounter cycles of exchanges involving a given number of particles).
Notice that in both 2D and 3D case, the pure solid and the $^3$He substitutional
solid has only one or two particle permutation cycles, 
while when an interstitial $^3$He or $^4He$ atom is introduced, 
it gives rise to permutations involving
up to a 10 atom chain, which is as long as the longest possible
distance in our lattice. This indicates that the
result may not be a finite-size effect.
\section{Discussion}
\label{discussion}
One of the main conclusions of the present paper is that the added interstitial
impurity in both 2D and 3D $^4$He becomes substitutional by creating 
a interstitial $^4$He defect; we believe that this result is firm and
it is not  subject to finite size effects. Furthermore, we find that
the effective mass of a $^3$He impurity atom in both 2D and 3D solid
$^4$He is large at $T=1$ K ($m^*/m \sim 5$) and at a lower temperature
of $500$ mK in 2D it becomes even larger ($m^*/m \sim 9$).   

In addition, we find that 
the above mentioned effect (i.e., the promotion of a $^4$He 
atom to the interstitial
band by the impurity) gives rise to a non-zero superfluid response 
and a significant enhancement of the OBDM at long-distances. 
This suggests that, provided that this effect persists when a finite
density of $^3$He impurities is present and, provided that such a 
metastable state can be created and
maintained, $^4$He solid with such impurities should be a supersolid.
However, this can not be established by the present calculation
done for a single impurity in a pure $^4$He solid
and it depends on a number of other factors. For example, while we have clearly
demonstrated that a single $^3$He impurity acts as a donor
of $^4$He atoms to the interstitial (``conduction'') band, the fate of 
these freed bosonic ``carriers'' is not certain when there is a finite
density of $^3$He impurities. In this case the created interstitial
$^4$He atoms can phase-separate in a similar way
as vacancies do\cite{Bon1}, or they may bind to existing defects, such as, 
dislocations, domain walls,  or grain boundaries
or even remain free.  
It is not clear that such interstitial
defects exist in the $^4$He solid caused by  $^3$He or
other impurities. This is an issue which could depend on the
process of the crystal growth\cite{smith,rudavskii,ganshin}. 

A 2D $^4$He solid only exists as films on substrates, such as 
on graphite. The phase diagrams of  first, second, third and fourth
layer of $^4$He on
graphite, as a function coverage, has been studied by PIMC simulation
\cite{Marlon}. The role of substrate corrugations, which is missing
from the present simulation of the ideal 2D $^4$He, is important
and the interplay of these substrate potential corrugations with the 
helium-helium interaction gives
rise to a wealth of interesting phases\cite{Marlon}.  
It is quite possible, however, that the main conclusion of the present
paper, that introducing an interstitial $^3$He impurity is solid $^4$He 
leads to the promotion of a $^4$He atom to an 
interstitial position while the $^3$He impurity becomes substitutional, may
remain valid even in the case of substrate corrugations.

The superfluid response which was calculated at 1 K and
is given in Table~\ref{table1} is very large considering the
fact that the calculation was done at such a high temperature.
This is a finite-size effect but at a much lower temperature the 
superfluid response is expected to be greater. A calculation of the superfluid
density at a significantly lower temperature requires much larger
computational time scales in order to be able to accurately sample it.
In the 3D case, the zero temperature condensate fraction
obtained as  the asymptotic value (infinite distance value) of the
off-diagonal OBDM at zero temperature, 
is much smaller by at least two orders of magnitude (as seen from 
Fig.\ref{fig10}). Therefore, as is well-known, there is a 
large factor relating the superfluid response and the actual condensate
fraction.
It is clear that introducing just a single impurity and taking the infinite
volume limit (or infinite area limit in 2D), the superfluid density
and the condensate fraction should vanish. It is interesting, however, 
the fact that the ratio of the values of both these two quantities 
to the impurity fraction (the impurity fraction is $1/N$, where $N$ 
is the total number of $^4$He atoms considered) is a number of order unity.
These reported results on the off-diagonal OBDM and superfluid
density, have only a qualitative value and one cannot draw
firm conclusions because of a) finite size effects and
b) they refer to the case of a single  $^3$He impurity or
single $^4$He interstitial. 

\section{Acknowledgments}
This work was partially supported by a NASA grant NAG3-2867 and the 
calculations were performed on the Florida State University 
High-Performance-Computing cluster. 



\begin{thebibliography}{99}

\bibitem{chan} E. Kim and M. H. W. Chan, Nature {\bf 427}, 225 (2004);
 Science {\bf 305}, 1941 (2004)
\bibitem{prokofev} N. Prokof'ev and B. Svistunov, Phys. Rev. Lett.
{\bf 94}, 155302 (2005).E. Burovski, et al., Phys. Rev. Lett.
{\bf 94}, 165301 (2005).
\bibitem{ceperley}
D. M. Ceperley and B. Bernu, Phys. Rev. Lett. {\bf 93}, 155303 (2004).
 B. K. Clark and D. M. Ceperley, Phys. Rev. Lett. {\bf 96}, 
105302 (2006). M. Boninsegni et al., Phys. Rev. Lett. {\bf 96}, 105301 (2006).
\bibitem{mas_triag}
M. Boninsegni and N. V. Prokof'ev, Phys. Rev. Lett. {\bf 95}, 237204 (2005).
\bibitem{worm}
M. Boninsegni, N.~V. Prokof'ev and B.~V. Svistunov,
	Phys. Rev. Lett. {\bf 96}, 070601 (2006);
	Phys. Rev. E {\bf 74}, 036701 (2006).
\bibitem{Bon1}  M. Boninsegni, A. B. Kuklov, L. Pollet,
N. V. Prokof'ev, B. V. Svitsunov, and M. Troyer, 
Phys. Rev. Lett. {\bf 97}, 0804101 (2006).
\bibitem{Pollet} L. Pollet, M. Boninsegni, A. B. Kuklov,
N. V. Prokof'ev, B. V. Svitsunov, and M. Troyer, 
Phys. Rev. Lett. {\bf 98}, 135301 (2007).
\bibitem{Bon2}  M. Boninsegni, A. B. Kuklov, L. Pollet,
N. V. Prokof'ev, B. V. Svitsunov, and M. Troyer, 
Phys. Rev. Lett. {\bf 99}, 035301 (2007).
\bibitem{pressure} L. Pollet, M. Boninsegni, A. B. Kuklov,
N. V. Prokof'ev, B. V. Svitsunov, and M. Troyer, 
Phys. Rev. Lett. {\bf 101}, 097202 (2008).
\bibitem{Dorsey} A. T. Dorsey, P. M. Goldbart, and J. Toner, Phys. Rev. 
Lett. {\bf 96}, 056301 (2006).
\bibitem{manousakis} E. Manousakis, Europhys. Lett. {\bf 78}, 36002 (2007).
\bibitem{Toner} J. Toner, Phys. Rev. Lett. {\bf 100}, 035302 (2008).
\bibitem{reppy}A.S.C.~Rittner and J.D.~Reppy, Phys. Rev. Lett., {\bf 97}, 
165301 (2006); {\bf 98}, 175302 (2007).
\bibitem{chan3}E. Kim and M. H. W. Chan, J. Low Temp. Phys. {\bf 138}, 
859 (2005). E. Kim J. S. Xia, J. T. West, X. Lin, A. C. Clark, 
and M. H. W. Chan, Phys. Rev. Lett
{\bf 100}, 065301 (2008).
\bibitem{goodkind} P.-C. Ho, I. P. Bindloss and J. M. Goodkind,
J. Low Temp. Phys. {\bf 109}, 409 (1997).
\bibitem{smith} A. Smith {\it et al.}, Phys. Rev. {\bf B 67}, 245314 (2003).
\bibitem{rudavskii}E. Rudavskii, A. {\it el al.} {\bf 121}, 713-718 (2000). 
\bibitem{ganshin} A. Ganshin, {\it et al.}, J. Low Temp. Phys.
{\bf 116}, 349 (1999). 
\bibitem{andreev} A. F. Andreev and I. M. Lifshitz, Sov. Phys. JETP {\bf 29},
1107 (1969). A. F. Andreev and A. E. Meierovich, Sov. Phys. JETP {\bf 40},
776 (1975).
\bibitem{andreev2} Andreev A.F., Sov.Phys. JETP {\bf 41}, 1170 (1976).
\bibitem{kagan}Y. Kagan, {\it Defects in Insulating Crystals}, 
p. 17, Springer Verlag, (Berlin, 1981). 
Y. Kagan and L.A. Maksimov, Sov. Phys. JETP {\bf 60}, 
201 (1984).
\bibitem{nmr}R. A. Guyer, R. C. Richardson and L. I. Zane, 
Rev. Mod. Phys. {\bf 43}, 532 (1971);
M. G. Richards {\it et al.}, J. Low. Temp. Phys. 
{\bf 24}, 1 (1976). M. G. Richards {\it et al.}, 
Phys. Rev. Lett. {\bf 34}, 1545 (1975);
V. A. Mikheev  et al., Solid State Comm. {\bf 48}, 
361 (1983);
A. R. Allen  et al., J. Low. Temp. Phys. {\bf 47}, 289 (1982).





\bibitem{cuervo}
J.~E. Cuervo, P.-N. Roy and M. Boninsegni,
	J. Chem. Phys. {\bf 122}, 114504 (2005).
\bibitem{aziz}
R.~A. Aziz {\it et~al.},
	J. Chem. Phys. {\bf 70}, 4330 (1979).
\bibitem{BC} M. Boninsegni and D. M. Ceperley,
Phys. Rev. Lett. {\bf 74}, 2288 (1995).

\bibitem{Marlon}M. Pierce and E. Manousakis,
Phys. Rev. Lett. {\bf 81}, 156 (1998); Phys. Rev. B {\bf 59}, 3802 (1999);
Phys. Rev. Lett. {\bf 83}, 5314 (1999); Phys. Rev. B {\bf 62}, 5228 (2000);
Phys. Rev. B {\bf 63}, 144524(2001).
\end{thebibliography}
\end{document}